\documentclass{JHEP3}
\usepackage{epsfig,multicol,bbm}

\renewcommand{\bar}[1]{\overline{#1}}

\newcommand{\beq}{\begin{equation}}
\newcommand{\eeq}{\end{equation}}

\def\kpc{{\rm kpc}}
\def\GHz{{\rm GHz}}
\def\GeV{{\rm GeV}}
\def\MeV{{\rm MeV}}
\def\cm{{\rm cm}}
\def\sec{{\rm s}}
\def\muGs{{\rm \mu Gs}}
\def\arcsec{{\rm arcsec}}

\def\microK{\mu{\rm K}}

\title{Dark matter annihilation and non-thermal Sunyaev-Zel'dovich effect:
II. dwarf spheroidal galaxy}

\author{Feng Huang$^{a,b}$,Xuelei Chen$^{a,c}$,Qiang Yuan$^{d,e}$,
Xiaojun Bi$^{d,c}$\\
$^a$  National Astronomical Observatories, Chinese Academy of Sciences,
Beijing 100012, China\\
$^b$ Department of Physics and Institute of Theoretical Physics and
Astrophysics, Xiamen University, Xiamen, Fujian 361005, China\\
$^c$ Center of High Energy Physics, Peking University, Beijing, 100871, China\\
$^d$ Key Laboratory of Particle Astrophysics,
Institute of High Energy Physics, Chinese Academy of Sciences,
Beijing 100049,China\\
$^e$ Graduate University of Chinese Academy of Sciences, Beijing 100049, China
}

\abstract{ We calculate the CMB temperature distortion due to the
energetic electrons and positrons produced by dark matter~(DM)
annihilation~(Sunyaev-Zel'dovich effect, $\rm SZ_{\rm DM}$) in
dwarf spheroidal galaxies~(dSphs). In the calculation we have
included two important effects which were previously ignored.
First we show that the $e^\pm$ with energy less than $\sim \GeV$,
which were neglected in previous calculation, could contribute a
significant fraction of the total signal. Secondly we also
consider the full effects of diffusion loss, which could
significantly reduce the density of $e^\pm$ at the center of cuspy
halos. For neutralinos, we find that detecting such kind of SZ
effect is beyond the capability of the current or even the next
generation experiments, which is consistent with the quantitative
description made by S.~Colafrancesco in \cite{draco_multi}. In the
case of light dark matter (LDM) the signal is much larger, but
even in this case it is only marginally detectable with the next
generation of experiment such as ALMA. We conclude that similar to
the case of galaxy clusters, in the dwarf galaxies the $\rm
SZ_{\rm DM}$ effect is not a strong probe of DM annihilations. }

\keywords{dark matter theory, Sunyaev-Zel'dovich effect, dwarf
spheroidal galaxies}

\begin{document}

\section{Introduction}

The physical nature of dark matter (DM) is a great unsolved problem
in modern cosmology. A major observational approach to this problem
is to look for possible signatures of DM annihilations. In many
theoretical models, the DM particles could annihilate and produce
$\gamma$-ray photons and other energetic particles such as electrons
and positrons. This is the case, for example, for the supersymmetric
DM model \cite{susyrev}, and the light dark matter (LDM) model
\cite{LDMtheory}. For some early investigations on this subject, see e.g.
\cite{Zeldovich:1980st,Rudaz:1987ry,Ellis:1988qp,Kamionkowski:1990ty,Fargion:1994me}.

Here we consider the energetic electrons and positrons produced by
such annihilations. One possible way to reveal or constrain the
presence of such energetic electrons and positrons is to look for
the Sunyaev-Zel'dovich (SZ) effect induced by them (SZ$_{\rm DM}$).
Energetic electrons or positrons could scatter with cosmic microwave
background (CMB) photons, producing a distortion in the CMB
spectrum. In clusters of galaxies, the virial temperature is
sufficiently high that this effect could be produced by thermal
electrons, which is the so called thermal SZ effect \cite{SZ1}. The
bulk movement of electrons with respect to the CMB also cause a
similar effect, which is dubbed the kinetic SZ effect \cite{SZ2}. In
galaxies, however, the virial temperature is too low for the thermal
electrons to produce significant thermal SZ effect, and the number
of free electrons is also too low to produce significant kinetic SZ
effect. However, non-thermal energetic electrons and positrons do
exist such as cosmic rays in galaxies, which may produce an
additional non-thermal SZ effect. In particular, DM annihilation may
supply some of such energetic particles. However, DM annihilation is
not necessarily the only source of cosmic ray particles. A number of
``normal'' astrophysical sites, such as supernova remnants, AGN
jets, pulsar winds etc., may also be partially responsible for
cosmic rays. This is why the dwarf spheroidal galaxies ~(dSphs) may
be of particular interest to us, as they are DM dominated and also
considered to be very inactive, hence observations are less affected
by background originating from normal astrophysical processes such
as star formation and AGN activity.

The SZ effect induced by a non-thermal distribution of charged
particles were studied in \cite{szn1,szn2}, and then applied to
clusters of galaxies~\cite{sz_cluster1,sz_cluster2,sz_cluster3}
and dSphs, where the Draco dwarf was taken as an prime
example~\cite{sz_draco}. These studies suggest that the DM-induced
SZ effect could be used to search for signatures of DM
annihilation. However, in a companion paper \cite{Yuan:2009yy}
(hereafter Paper I), we calculated the DM-induced SZ effect for
clusters. We showed that contrary to previous claims, the DM
annihilation induced SZ effect in galaxy clusters is not as large
as previously work~\cite{sz_draco}, and there is little hope to
detect it in the foreseeable future, which confirm the
quantitative conclusion in \cite{draco_multi}. Similar conclusion
is also derived in a recent study \cite{2010JCAP...02..005L}. In
the present paper, we generalize our earlier work to the case of
dSphs, for two DM candidates: the neutralino and the LDM.

The rest of the paper is organized as follows: in Section 2, we
discuss our method of calculation. We first briefly review the
equations for SZ calculation with a given electron distribution.
Then we discuss the density distribution of the DM in dSphs.
Lastly the production of electron s and positrons from DM
annihilation and propagation of these $e^{\pm}$ are discussed.
This differs somewhat from the case of clusters, as diffusion is
more significant. In Section 3, we present the results of our
numerical calculation. Finally we discuss our results and conclude
in Section 4.

\section{Method of Calculation}

\subsection{The nonthermal SZ Effect}

For the calculation of the SZ effect, our method here is the same
as in Paper I (see also \cite{szn1}). The spectral distortion of
CMB after travelling through a population of electrons or
positrons \footnote{As all the calculations are the same for
electrons and positrons, in this paper, the subscript $e$
represents both $e^+$ and $e^-$} is given by
\begin{equation}
\frac{\Delta
T(x,\theta)}{T_0}=\frac{(e^x-1)^2}{x^4e^x}g(x)y(\theta),
\label{temperature}
\end{equation}
where $x=h\nu/kT_0$ is the dimensionless frequency of CMB photon,
$T_0= 2.725$ K is the undistorted CMB temperature, $g(x)$ is the
spectral distortion function, and $y(\theta)$ is the Comptonization
parameter for angle separation $\theta$ from the center. The
spectral distortion function is given by
\begin{equation}
g(x)=\frac{m_ec^2}{\langle kT_e\rangle}\left[\int
i_0(xe^{-s})P_1(s){\rm d}s-i_0(x)\right],
\end{equation}
where $i_0(x)=x^3/(e^x-1)$ is the Planckian distribution of CMB
photons, $s=\ln(\nu^{\prime}/\nu)$ is the frequency shift of one
photon after one scattering with $e^{\pm}$. Here $P_1(s)$ is the
frequency shift probability distribution after one scattering,
$$P_1(s)=\int f_e(k)P_s(s,k){\rm d}k,$$
where $k$ is the dimensionless momentum (momentum in units of $m_e
c$), $P_s(s,k)$ is the probability of a photon to have a frequency
shift $s$ when colliding with an electron with momentum $k$,
$f_e(k)$ is the normalized momentum spectrum of electrons
$$f_e(k)=\frac{1}{n_e}\frac{{\rm d}n_e}{{\rm d}E_e}\frac{{\rm d}E_e}{{\rm d}k},$$
with $\frac{{\rm d}n_e}{{\rm d}E_e}(E_e,r)$ the number density of
$e^{\pm}$ per unit energy interval, which will be discussed in
details in the next section when the dSphs and DM candidate are
specified. The number density of non-thermal electrons is
\begin{equation}
n_e(r)=\int_{E_{\rm min}}^{E_{\rm max}}\frac{{\rm d}n_e}{{\rm
d}E_e}\left(E_e,r\right){\rm d}E_e, \label{ne}
\end{equation}
with $E_{\rm max}$ and $E_{\rm min}$ the maximum energy and minimum
energy of electrons produced by DM annihilation. The effective
pressure of the non-thermal electrons can be defined as (see e.g.
\cite{CR1,CR2})
\begin{eqnarray}
P_{e}= n_e\int^{k_{\rm max}}_{k_{\rm min}} f_e(k) \left(\frac{k^2}{3
\sqrt{1+k^2}}\right)m_e c^2 {\rm d}k, \label{Pe}
\end{eqnarray}
with $k_{\rm max}=\sqrt{(E_{\rm max}/m_e)^2-1}$, and $k_{\rm
min}=\sqrt{(E_{\rm min}/m_e)^2-1}$. In the relativistic limit, this
definition approaches to the usual approximation
\begin{equation}
P_{e}\approx \frac{1}{3} \langle E\rangle, \qquad \langle
E\rangle=\int^{k_{\rm max}}_{k_{\rm min}} f_e(k)
\left(\sqrt{1+k^2}-1\right)m_e\,{\rm d}p , \label{eq:E}
\end{equation}
in which $\langle E\rangle$ is the averaged kinetic energy. When DM
annihilation is the only source of $e^{\pm}$, the upper limit of the
integrals are set to be $E_{\rm max}=m_{\rm DM}$ and $k_{\rm
max}=\sqrt{(m_{\rm DM}/m_e)^2-1}$. In some analyses~\cite{sz_draco},
$E_{\rm min}$ is set to be $0.01~m_{\rm DM}$, which was considered
to be a good approximation as it is small enough comparing with the
peak of the energy spectrum of continuous electrons~($E_{\rm
peak}=1/20~m_{\rm DM}$) as shown in \cite{Kamionkowski:1990ty}.
However, for the non-thermal SZ effect in dwarf galaxies, the
contribution of electrons with $E<0.01 m_{\rm DM}$ to the final
signal is still significant. Therefore, we use Eq. (\ref{Pe}) which
gives a more accurate expression for the contribution of
trans-relativistic electrons and set $E_{\rm min}=m_e$ as no
artificial cut off is necessary. Note however that even though this
expression itself is accurate, there could still be significant
error in the expression of $f_e(k)$ used for calculation at the
lowest energies, see the discussion in \S 2.4 on $e^+e^-$ production
from DM annihilations.

The contribution to Comptonization parameter by DM annihilation is
given by the line-of-sight integral of effective gas pressure
through the $e^\pm$ cloud:
\begin{equation}
y=\frac{\sigma_{T}}{m_{e}c^{2}}\int P_e \mathrm{d}l,
\label{eq:yparam1}
\end{equation}
where $\sigma_{T}$ is the Thomson cross section.

The observed effective temperature change due to SZ$_{\rm DM}$ is an
average of the temperature change within a beam
\begin{equation}
\Delta T_{obs}({\bf n},\theta)=\int B_{{\bf n}}({\bf n'},\theta)
\Delta T({\bf n'}) {\rm d}{\bf n'},
\end{equation}
where $B_{{\bf n}}({\bf n'},\theta)$ is the beam profile in the
direction ${\bf n'}$ for a beam centered in the direction ${\bf n}$
with beam size $\theta$. For illustration, we have made our
calculation with a tophat beam profile and different beam widths.

\subsection{dark matter distribution}

In dSphs, DM dominates these systems both in the inner parts and
outskirts. Similar as in Paper I, we consider the following three
types of DM density profiles:
\begin{eqnarray}
\rho(r)&=&\frac{\rho_{\rm s}}{(1+r/r_s)[1+(r/r_s)^2]}\ ({\rm
hereafter\ B95,\ Ref.} \cite{Burkert}),\\
\rho(r)&=&\frac{\rho_{\rm s}}{(r/r_s)(1+r/r_s)^2}\ \ \ \ \ \ \ \
({\rm hereafter\ NFW,\ Ref.} \cite{NFW}),\\
\rho(r)&=&\frac{\rho_{\rm s}}{(r/r_s)^{1.5}[1+(r/r_s)^{1.5}]}\ ({\rm
hereafter\ M99,\ Ref.} \cite{Moore}) .
\end{eqnarray}
They show similar behaviors ($\sim r^{-3}$) at large radii, but
differs significantly near the center of the halo. We employ the
B95, NFW and M99 profiles to represent the non-cuspy, moderately
cuspy and strongly cuspy profiles of DM halos respectively.

The density profile is truncated at a small radius where we assume
the annihilating rate matches the in-falling rate of
DM~\cite{1992PhLB..294..221B}. Within this radius, the density is
kept at a constant value as
\begin{equation}
\rho_{\rm max}= 3\times10^{18}\frac{m_{\rm DM}}{100\,{\rm
GeV}}\frac{10^{-26}{\rm cm}^2{\rm s}^{-1}}{\langle\sigma
v\rangle}\,{\rm M}_{\odot} {\rm kpc}^{-3}\ .
\end{equation}

The density distribution of DM could in principle be determined from
the velocity dispersion of stars. However, the uncertainty is large,
and in the literature the required data is not always complete or
easy to use. Parameters such as the characteristic density $\rho_s$
and radius $r_s$ are chosen to be set by adopting specific density
profile. Another generally used parameters are the virial mass $\rm
M_{\rm vir}$ and concentration parameters $c_{\rm vir}$. The virial
radius of a DM halo is defined as
\begin{equation}
r_{\rm vir}=\left(\frac{\rm M_{\rm vir}}{(4\pi/3)\Delta
\rho_c}\right)^{1/3}\,
\end{equation}
where $\Delta$ is the overdensity, and $\rho_{\rm c}\approx 139$
M$_{\odot}$ kpc$^{-3}$ is the critical density of the universe.
For the $\Lambda$CDM universe, $\Delta\approx 18\pi^2+82x-39x^2$ with
$$x=\Omega_M(z)-1=-\frac{\Omega_{\Lambda}}{\Omega_M(1+z)^3+
\Omega_{\Lambda}}$$
is found to be a good approximation
\cite{1998ApJ...495...80B}. The concentration parameter $c_{\rm
vir}$ is defined as
\begin{equation}
c_{\rm vir}=\frac{r_{\rm vir}}{r_{-2}}, \label{cv}
\end{equation}
where $r_{-2}$ refers to the radius at which $\frac{{\rm d} \left(
r^2\rho \right)}{dr} |_{r=r_{-2}}=0$. The concentration parameter
$c_{\rm vir}$ relates $r_{\rm vir}$ and the density profile
parameter as \cite{2001MNRAS.321..559B}
\begin{equation}
r_{\rm s}^{\rm B95}=\frac{r_{\rm vir}}{1.52c_{\rm vir}},\ r_{\rm
s}^{\rm NFW}=\frac{r_{\rm vir}}{c_{\rm vir}},\ r_{\rm s}^{\rm
M99}=\frac{r_{\rm vir}}{0.63c_{\rm vir}}. \label{rs}
\end{equation}
Simulations show that $c_{\rm vir}$ and $M_{\rm vir}$ are often
correlated. Here we use a power-law relation between $c_{\rm vir}$
and $M_{\rm vir}$ as suggested in Ref.~\cite{2004A&A...416..853D}
\begin{equation}
c(M_{\rm vir})=c_0\times\left(\frac{M_{\rm vir }}{10^{14}h^{-1}{\rm
M}_{\odot}}\right)^{ \alpha},
\end{equation}
where $c_0=9.6$ and $\alpha=-0.1$ for a $\Lambda$CDM cosmology. With
this assumption, the DM distribution is fixed once $M_{\rm vir}$ is
given.

The halo of the dSphs would, however, be disrupted by the tidal
force of the host galaxy (Milky Way), as a result part of the DM at
the outer edge of the halo would be removed. To account for this
effect, we calculate the tidal radius, which is given by the Roche
criterion~\cite{Evans2004}
\begin{equation}
\frac{M_{\rm dSph}(r_t)}{r_t^3}=\frac{M_{\rm MW}(\rm D-r_t)}{(\rm
D-r_t)^3},
\end{equation}
where D is the distance between the Galactic center and the dSph,
$r_t$ is the tidal radius of the dSph and $M(r)$ is the mass
inside radius $r$. The results depend on the choice of the profile
for Milky Way halo. Here we adopted the isothermal power law
model~\cite{CPLmodel} for the Milky Way
\begin{equation}
\rho_{\rm iso}=\frac{v_a^2}{4\pi G}
\frac{3r_s^2+r^2}{(r_s^2+r^2)^2}
\end{equation}
with $r_s=10$ kpc and $v_a=220$ km~s$^{-1}$ as in
\cite{sz_draco,Evans:2003sc}.

In Table~\ref{15dSphs}, we list the parameters for a sample of 7
nearby dSphs, which were obtained by fitting the line-of-sight
velocity dispersion profile of the stars~\cite{7dSphs}. NFW
profile was assumed when deriving these parameters. Strictly
speaking, these parameters should not be used for other halo
profiles. To be self-consistent, one should re-estimate the virial
mass with different halo profiles. However, since there are many
uncertainties in both the observation and theoretical modeling,
and the results of this paper should be regarded as an
order-of-magnitude estimate, we will just use the same virial
mass for M99 and B95 profiles but re-evaluate the halo parameters.

\begin{table}
\begin{center}
\caption{ \label{15dSphs} The dSphs parameters used in this paper.
NFW profile of dSph is assumed when calculating $r_s$, $\rho_s$
and $r_t$. The distance and virial mass data are mostly taken from
Ref.~\cite{7dSphs}, other parameters are calculated as outlined
below. For Ursa Minor, which was not included in
Ref.~\cite{7dSphs}, we take the mass as to be the same as that of
Draco.}
\begin{tabular}{lccccc}
\hline \hline Name &$\frac{\rm D}{[\rm kpc]}$& $\frac{\rm M_{\rm
vir}}{[10^8\rm M_{\odot}]}$&$\frac{\rho_s}{[10^8\rm M_{\odot}/\rm
kpc^3]}$&$\frac{r_s}{[\rm kpc]}$&$\frac{r_t}{[\rm kpc]}$\\
\hline
Draco & $80$&$40$&0.82&1.2&9.9\\
LeoI & $250$ &$10$&1.2&0.64&16.7\\
Fornax & $138$ &$10$&1.2&0.64&10.3\\
LeoII & $205$ &$4$&1.5&0.43&10.6\\
Carina & $101$ &$2$&1.8&0.32&4.8\\
Sculptor & $79$ &$10$&1.2&0.64&6.5\\
Sextans & $86$ &$3$&1.6&0.38&4.8\\
Ursa Minor& $66$&$40$&0.82&1.2&8.6\\
\hline \hline
\end{tabular}
\label{tab:sample}
\end{center}

\end{table}

The nearby and relatively large Draco dwarf represents one of the
best candidates for searching for DM in dSphs, and it has been
studied extensively with many observations at different wavelengths.
Since the basic properties of the SZ effects in different dSphs are
qualitatively similar to each other, we will take the Draco dwarf as
the prime example in the following discussions.

\subsection{$e^\pm$ production from DM annihilation}
As the SZ effect produced by positrons and electrons can not be
distinguished, here we use $e$ to refer both $e^+$ and $e^-$. The
$e^\pm$ source function from DM annihilation can be written as
\begin{equation}
Q_e(E,r) = \frac{\langle\sigma v\rangle}{2\,m_{\rm DM}^2} \sum_f
\frac{{\rm d}N_{e}^f}{{\rm d}E_e}(E) B_f\; \rho^2(r)\;,
\label{eq:epmsource}
\end{equation}
where $\langle\sigma v\rangle$ is the velocity-weighted annihilation
cross section, $m_{\rm DM}$ is the DM particle mass, $\rho(r)$ is
energy density of DM, and ${\rm d}N_e/{\rm d}E_e$ is the number of
electrons produced per annihilation per energy interval. In the
following, we discuss two types of DM particles: neutralino in the
supersymmetric model, and the LDM.

{\it Neutralino:} We consider a neutralino with the typical values
of parameters $\langle\sigma v\rangle=3\times10^{26}~{\rm cm}^3~{\rm
s}^{-1}$ and $m_{\rm DM}=100~\GeV$. The direct channel to $e^+e^-$
is generally suppressed for neutralino, so electrons are in most
cases produced from the cascades of the annihilation final-state
particles such as heavy leptons, quarks and gauge bosons
\cite{susyrev}. The spectra of electrons can be different from each
other for different annihilation modes. We use the package DarkSUSY
\cite{DarkSUSY} to calculate the final-state spectra of electrons.
Our fiducial annihilation mode is assumed to $W^+W^-$.

\FIGURE{
\includegraphics[scale=0.5]{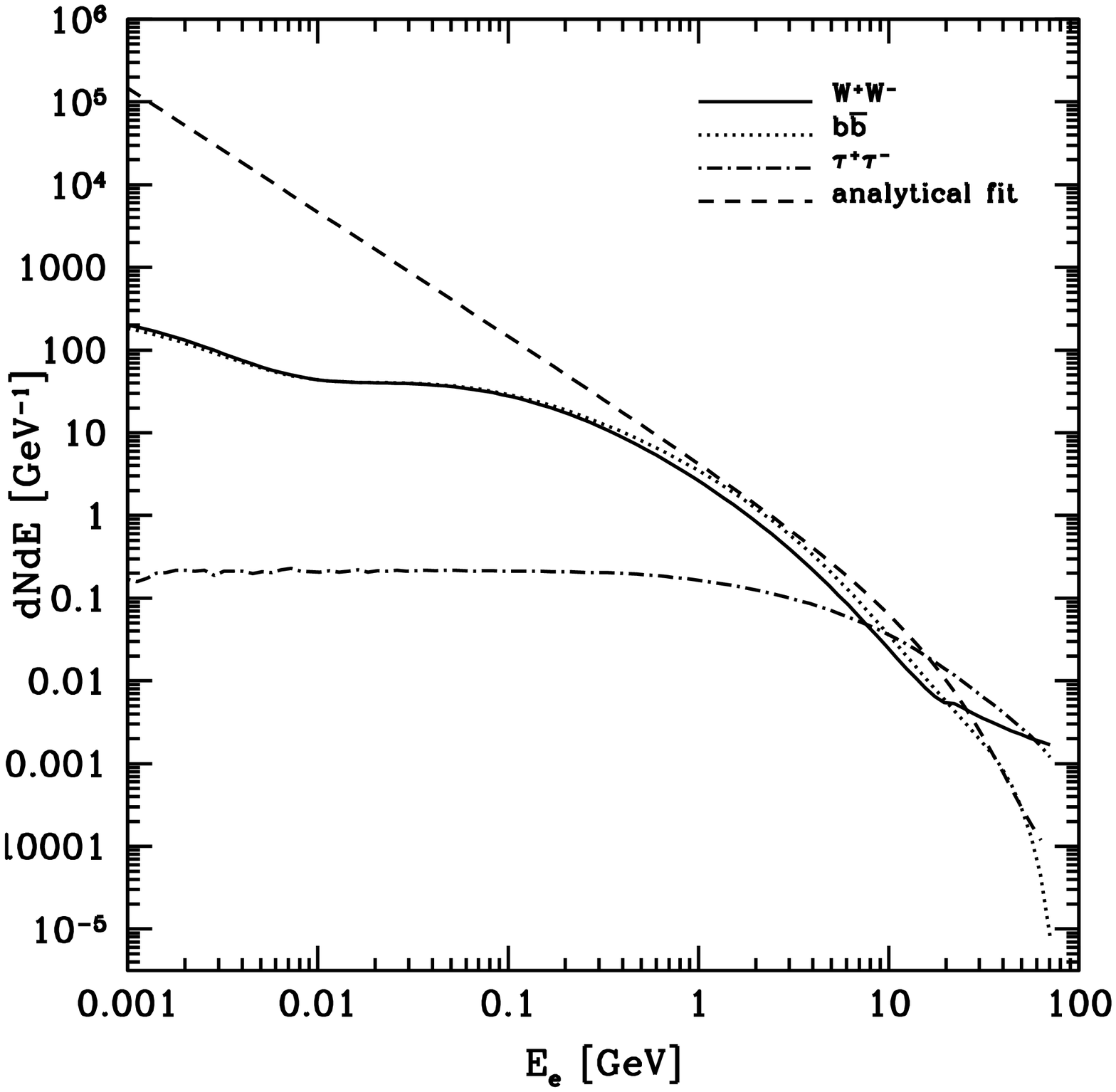}
\caption{ Electron yield spectra ${\rm d}N/{\rm d}E$ for
$W^+W^-$~(solid curve), $b\bar{b}$~(dotted curve) and
$\tau^+\tau^-$~(dot-dashed curve) annihilation modes for neutralino
with mass $m_{\chi}=100$ GeV. The dashed curve refers to the
analytical fit adopted in \cite{sz_draco}.} \label{dNdE} }

In Fig.~\ref{dNdE}, we plot the resulting electron energy spectrum
for several annihilation modes. In the same figure, we also plot the
parametrized fit of spectra used in \cite{sz_draco} for comparison.
As we can see from the figure, the parametrization fits the
DarkSUSY spectrum well above $1~\GeV$. However below $0.1~\GeV$ this
analytical form deviates significantly from the DarkSUSY results.


{\it LDM:} Inspired by the excess of $511~\rm keV$ lines at the
Galactic center observed by SPI/INTEGRAL \cite{511keV1,511keV2}, it
was proposed that a light DM particle with mass $1-100~\MeV$ may be
responsible for the data \cite{boehm0309686}. To allow the
production of positrons through its annihilation, one needs $m_{\rm
DM}>m_e$. The mass of LDM should also be less than $\sim 100 \MeV$
to avoid producing too many $\gamma$-ray photons from $\pi$ final
state. The model has since been further constrained by a number of
more detailed considerations on its
phenomenology~\cite{LDM1,LDM2,Beacom:2005qv,LDM3,LDM4}. For example,
a small fraction of the energetic positrons produced in the
annihilation could be directly annihilated into energetic
$\gamma$-rays (in-flight annihilation), and the limit on
$\gamma$-rays yields stringent limits of $m_{\rm DM}<3 \MeV$
\cite{Beacom:2005qv} or $m_{\rm DM}<7.5 \MeV$ \cite{LDM3}, depending
on the ionization state of the interstellar medium in the Galactic
center. A comparable bound of $m_{\rm DM}<7 \MeV$ was derived using
CMB data by considering the effect of LDM annihilation during the
recombination history \cite{LDM4}.

Here, as in Paper I, we assume the mass of the LDM is 5~MeV, which
is consistent with the limits obtained in \cite{LDM3,LDM4}. It is
slightly beyond the constraint of \cite{Beacom:2005qv}, but note
that the SZ effect here would be stronger for smaller DM mass, so
our choice is actually conservative. The cross section derived
from the flux of the 511 keV line emission from the bright
Galactic bulge region is $\langle\sigma v\rangle_{\rm LDM} \sim
10^{30}(m_{\rm DM}/\rm MeV)^2~\cm^3~s^{-1}$. Thus for our mass
choice of LDM we have $\langle\sigma v\rangle=2.5
\times10^{-29}~\cm^3~{\rm s}^{-1}$. The recent observations of the
spatial morphology of the line emission by INTEGRAL
\cite{Weidenspointner:2008zz} found a significant asymmetry in the
disk emission with a resemblance to the observed distribution of
low-mass X-ray binaries in the hard state, which indicates that
these X-ray binaries might be the main sources. However, it is
worthwhile to note that the large bulge-to-disk ratio~($3\sim9$)
is not easily achieved even with the assumption that a large
fraction of the disk positrons is transported via the regular
magnetic field of the Galaxy into the bulge and annihilate there,
so the LDM model can not be excluded yet. Our assumption about the
annihilating cross section may be regarded as the upper limit of
LDM contribution to the observed 511 keV line. We assume the
produced $e^\pm$ have a monochrome spectra ${\rm d}N_e/{\rm
d}E_e=\delta(E-m_{\rm DM})$ in this case.

\subsection{$e^\pm$ propagation in dwarf galaxies}

DM annihilation injects electrons and positrons in the galaxy halo
at a constant rate. The propagation of these charged particles in
the tangled magnetic field can be modeled by diffusion. They also
lose energy by radiation during this process. As a result, the
$e^\pm$ spectrum satisfies the following transport equation
\cite{book}:
\begin{equation}
  \frac{\partial}{\partial t}\frac{{\rm d}n_e}{{\rm d}E_e} =
  \nabla \left[ D(r,E) \nabla\frac{{\rm d}n_e}{{\rm d}E_e}\right] +
  \frac{\partial}{\partial E} \left[ b(r,E) \frac{{\rm d}n_e}{{\rm d}E_e}
  \right]+Q_e(r,E) \;,
\label{diffeq}
\end{equation}
where ${\rm d}n_{e}/{\rm d}E_{e}$ is the number density of
$e^{\pm}$ per unit energy interval, $D(E,r)$ is the diffusion
coefficient, $b(E,r)=-{\rm d}E_e/{\rm d}t$ represents the energy
loss rate, and $Q_e(E,r)$ is the source function. For simplicity,
we assume that $D$ and $b$ are independent of spatial location.
The diffusion coefficient $D(E)$ is assumed to have a power law
dependence on energy $E$ and magnetic field $B$: $D(E) = D_0
\left(E/B\right)^{\delta}$, with $\delta=1/3$ \cite{sz_draco},
although it is not clear to what extent this relation could be
extrapolated. Due to the much smaller scale of uniformity of the
magnetic field in dwarf galaxies, we set $D_0 = 3.1 \times
10^{26}$~cm$^{2}$~s$^{-1}$, which is lower than the value in the
Milky Way \cite{draco_multi,dwarf_xray}. We also set
$B_{\mu}=1~\muGs $ as our fiducial value of magnetic field.

The energy loss rate is given by $b(E)=b_{\rm ICS}(E)+b_{\rm
syn}(E)+b_{\rm ion}(E)$, i.e.
\begin{equation}
\frac{b(E)}{10^{-17}\rm GeV~s^{-1}}= 2.5\times\left(\frac{\beta
E}{1\rm GeV}\right)^2+0.25\times \left(\frac{B_{\mu}}{1~\mu
G}\frac{\beta E}{1\rm GeV}\right)^2+2\times\frac{N_{\rm H}}{1\rm
cm^{-3}}[\ln(\Gamma)+6.6],
\end{equation}
where $N_{\rm H}\simeq 1.3\times 10^{-6} \cm^{-3}$ is the number
density of neutral gas in dSphs, $\Gamma$ and $\beta$ are the
Lorentz factor and the velocity of $e^{\pm}$ respectively.

\FIGURE{
\includegraphics[scale=0.43]{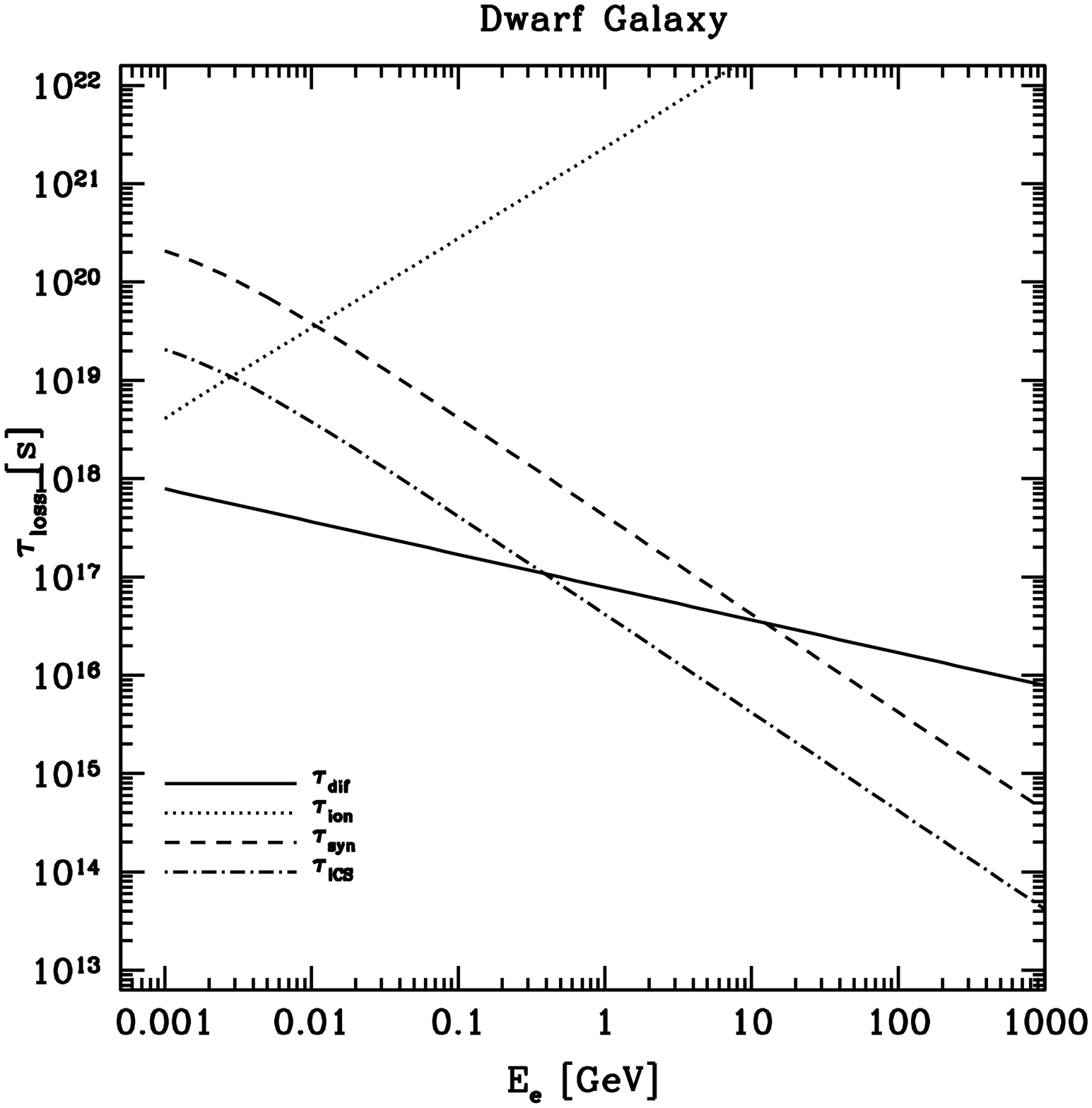}
\caption{Comparison of the time scales for various energy loss
mechanisms and diffusion in a dSph galaxy with size $r_h=1.6$~kpc.
$B=1\muGs$ and $N_{\rm H}=1.3\times 10^{-6} \cm^{-3}$ are
assumed.}\label{tloss} }

The characteristic time scales for different energy loss
mechanisms and spatial diffusion are plotted in Fig. \ref{tloss}.
The energy loss time-scale is defined as $\tau_{\rm loss} =
E/b(E)$, and the diffusion time-scale is $\tau_{\rm dif}\approx
r^2_h/ D(E)$. We clearly see that the ionization loss has
negligible effects, except below MeV energies. However, as the
size of the dwarf galaxy is small, spatial diffusion could
dominate at this energy range. Synchrotron emission and inverse
Compton scattering (ICS) off the CMB photons are important for
high energy electrons.

The transport equation (\ref{diffeq}) could be solved with the
Green's function method as described in \cite{draco_multi}. For
time-independent source function, the equilibrium solution has the
form
\begin{equation}
\frac{{\rm d}n_e}{{\rm d}E}\left(r,E \right) = \frac{1}{b(E)}
\int_E^{m_{\rm DM}} {\rm d}E' \; \widehat{G}\left(r,\Delta v\right)
Q_e(r,E'). \label{eq:full2}
\end{equation}
where $\Delta v(E,E')=\int_E^{E'} {\rm
d}\epsilon\,D(\epsilon)/b(\epsilon)$ is the mean scale of diffusion
covered by an electron while losing energy from $E^\prime$ to $E$.
For a spherical symmetric system with free escape boundary condition
at $r_h$ (i.e. assuming the magnetic field is so weak that beyond
this region the electrons are not confined), the Green's function is
given by
\begin{equation}
\widehat{G}\left(r, \Delta v\right) = \frac{1}{\sqrt{4\pi\Delta
v}} \sum_{n=-\infty}^{+\infty} (-1)^n \int_0^{r_h} {\rm d}r'
\frac{r'^2 \rho^2(r')}{rr_n' \rho^2(r)}
\left[\exp{\left(-\frac{(r-r_n')^2}{4\,\Delta v}\right)}-
\exp{\left(-\frac{(r+r_n')^2}{4\,\Delta v}\right)}\right]
\;,\label{eq:rescaling}
\end{equation}
where $r_n' = (-1)^n r' + 2 n r_h$ is the location of $n$th
``charge'' image for $r_h$ (c.f. Ref.~\cite{sz_cluster1}). The
value of $r_h$ is generally adopted as twice of the radius of the
stellar component, typically a few kpc for local dSphs.

\FIGURE{
\includegraphics[scale=0.45]{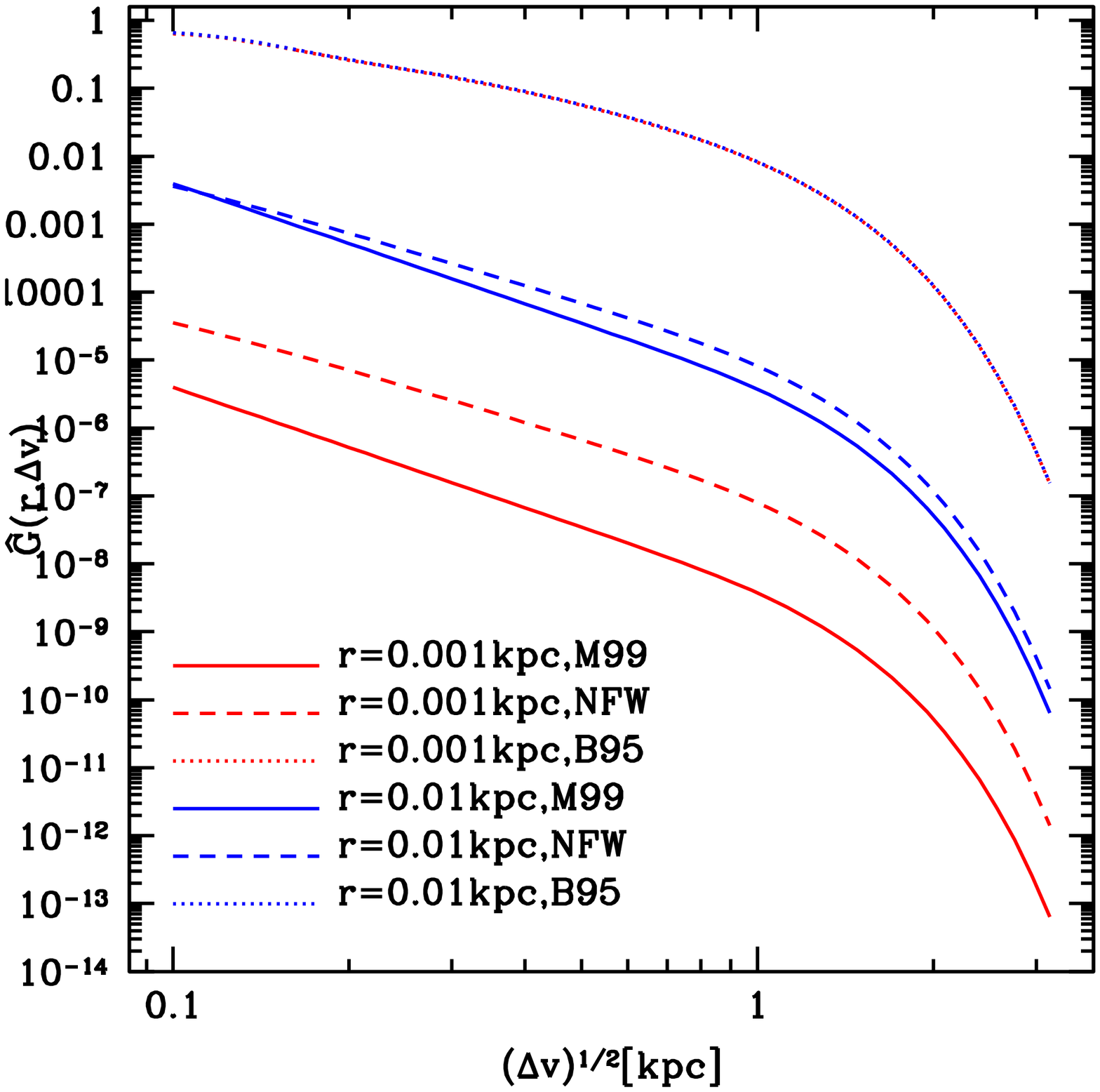}
\caption{The Green's function $\widehat{G}$ varies with respect to
$\sqrt{\Delta v}$ for two different radii $r=0.001~\rm kpc$ and
$r=0.1~\rm kpc$ and for three density profiles: B95, NFW and M99.
\label{Green}} }

The behavior of the Green's function $\widehat{G}\left(r, \Delta
v\right)$ is shown in Fig. \ref{Green}. We plot $\widehat{G}$ as a
function of $\sqrt{\Delta v}$ for two different radii and for the
three density profiles described in Sec. 2.2. As expected, the
Green's function decreases as the diffusion length $\sqrt{\Delta v}$
increases. Near the center of the galaxy, however, we find
$\widehat{G} \ll 1$ as represented by the $r=0.001~\kpc$ curves for
the M99 and NFW profiles, indicating that the diffusion effect is
significant and it will affect the propagated spectrum. We can also
see that for more cuspy density profile, the decrement in
$\widehat{G}$ is also larger.

When spatial diffusion can be neglected, $\widehat{G} \sim 1$, we
obtain the solution for no diffusion or {\it in situ} energy loss:
\begin{equation}
\left.\frac{{\rm d}n_e}{{\rm d}E}\left(r,E \right)\right|_{\rm
nodif}=\frac{1}{b(E)}\int_E^{m_{\rm DM}}{\rm d} E^{\prime}Q_e(
r,E^{\prime}), \label{equili}
\end{equation}
The spatial distribution of $e^{\pm}$ then traces the source
function, which is proportional to the density square of DM.
However, for dSphs this approximation is generally invalid.

\section{Results}

\subsection{Distribution of electrons}

To take the diffusion effect into account, Ref.~\cite{sz_cluster1}
also proposed an analytic approximate solution of the transport
equation
 \beq
 \frac{{\rm d}n_e}{{\rm d}E}\left(r,E \right) \approx [Q_e(E,r) \tau_{loss}]
 \times {V_{s} \over V_s + V_o} \times {\tau_{D} \over \tau_{D}+ \tau_{loss}},
 \label{eq.solution.qualitative}
 \eeq
where $V_s \propto r^3_h$ is the volume occupied by the DM
source\footnote{As the galactic magnetic field has a limited extent,
here we define $r_h$ as the radius of diffusion zone instead of the
tidal radius of the dark halo.}, and $V_o \propto \lambda^3(E)$ is
the volume occupied by a diffusing electron which travels a distance
$\lambda(E) \approx [D(E) \times \tau_{loss}(E)]^{1/2}$ before
losing a significant fraction of its initial energy. This changes
the total number of electrons, but does not change the shape of its
density profile. Here we call this model as the ``approximate
solution''.

\FIGURE{
\includegraphics[scale=0.45]{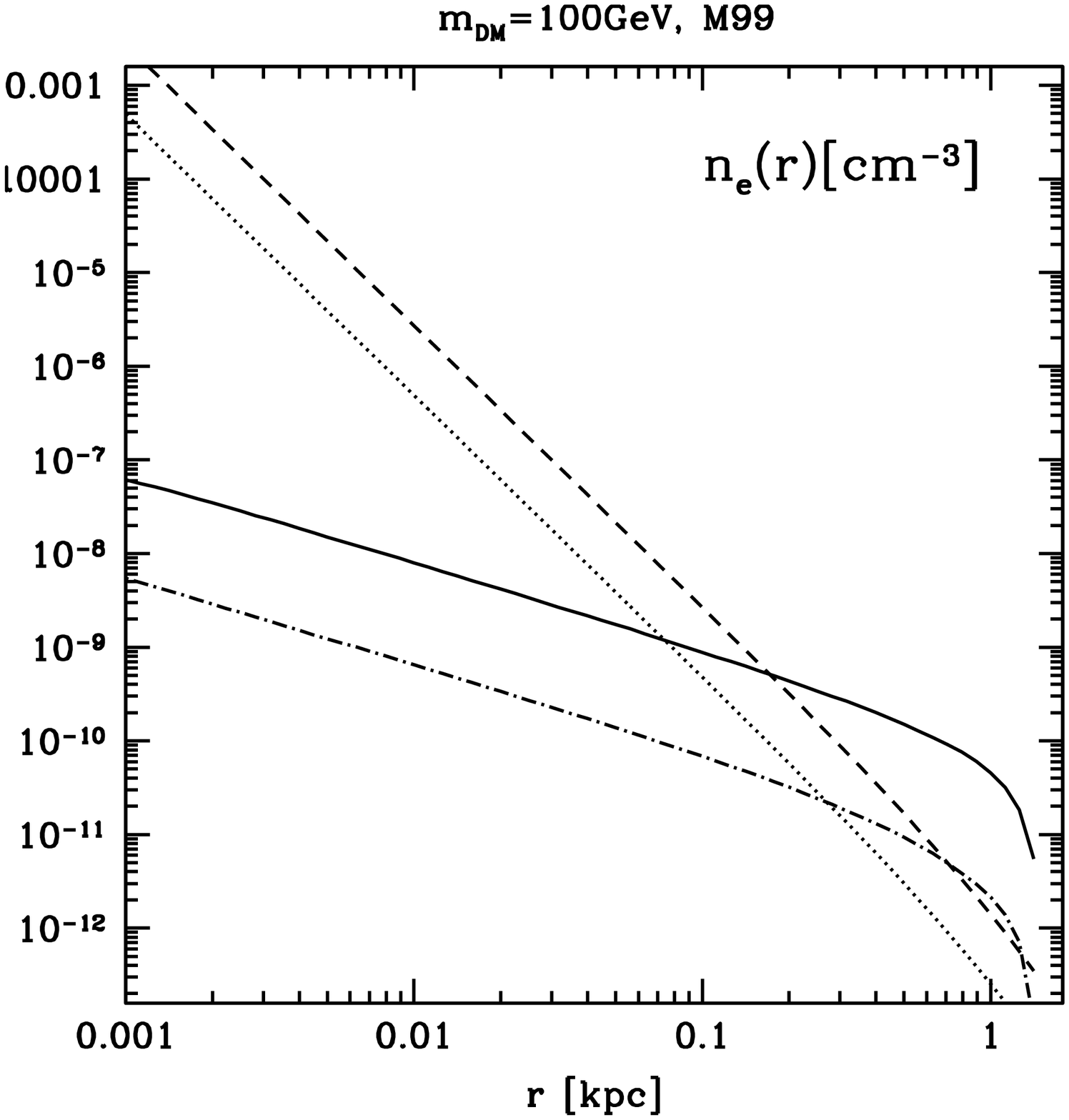}
\caption{The $e^\pm$ density $n_e(r)$ from neutralino annihilations
in the Draco dSph. The solid and dot-dashed curves are our full
solutions with $E_{\rm min}=m_e$ and $0.01m_{\rm DM}$ respectively.
The dotted curve is the result of the ``approximate solution'', and
the dashed curve is the result of the ``no-diffusion solution'' as
described in Eq.~(\ref{eq:nodif}). \label{WIMPne}} }

We plot in Fig. \ref{WIMPne} the total number density distribution
of $e^\pm$ produced by neutralino annihilation for the case of
dSph Draco with an assumed M99 profile. Our full solution obtained
with the Green's function method is shown as the solid curve,
while the ``approximate solution'' is shown as the dotted one. The
dot-dashed curve below the solid one is the result obtained with
$E_{\rm min}=0.01 m_{\rm DM}$. We see that the non-thermal
electrons with $E<0.01m_{\rm DM}$ still make up a significant
fraction of the total density. Previous analysis on SZ$_{\rm DM}$
in dwarf galaxies in Ref.~\cite{sz_draco} predicted
$\microK$-level temperature distortion within $1~\arcsec$ for
strongly cusped (M99) dark halo. They employed the approximation
\begin{equation}
n_e(r)=\int_{0.01m_{\rm DM}}^{m_{\rm DM}} \left.\frac{{\rm
d}n_e}{{\rm d}E}(r,E)\right|_{\rm nodif} {\rm d}E. \label{eq:nodif}
\end{equation}
This ``no-diffusion solution'' is shown as the dashed line. It is
apparent that there are significant differences between the full
solution and the simplified solution or the approximate solution.
The $e^\pm$ density profile of our full solution has a
significantly shallower slope than the ``approximate solution''.
Thus in the inner region, the full solution predicts much lower
electron number density than either the ``no-diffusion solution''
or the ``approximate solution''. We can understand this by noting
that diffusion could remove electrons in the inner region and
spread them in a larger region. Although the ``no-diffusion
solution'' attempts to account for the effect of diffusion by
reducing the total number density of electrons, it reduces the
electron density by the same factor at all radii. Actually the
inner region will be affected much more significant than the outer
regions. Also, for more cuspy halos, the scale over which
$\rho^2(r)$ has a significant gradient is smaller, resulting in
more remarkable loss of electrons due to diffusion.

\FIGURE{
\includegraphics[scale=0.45]{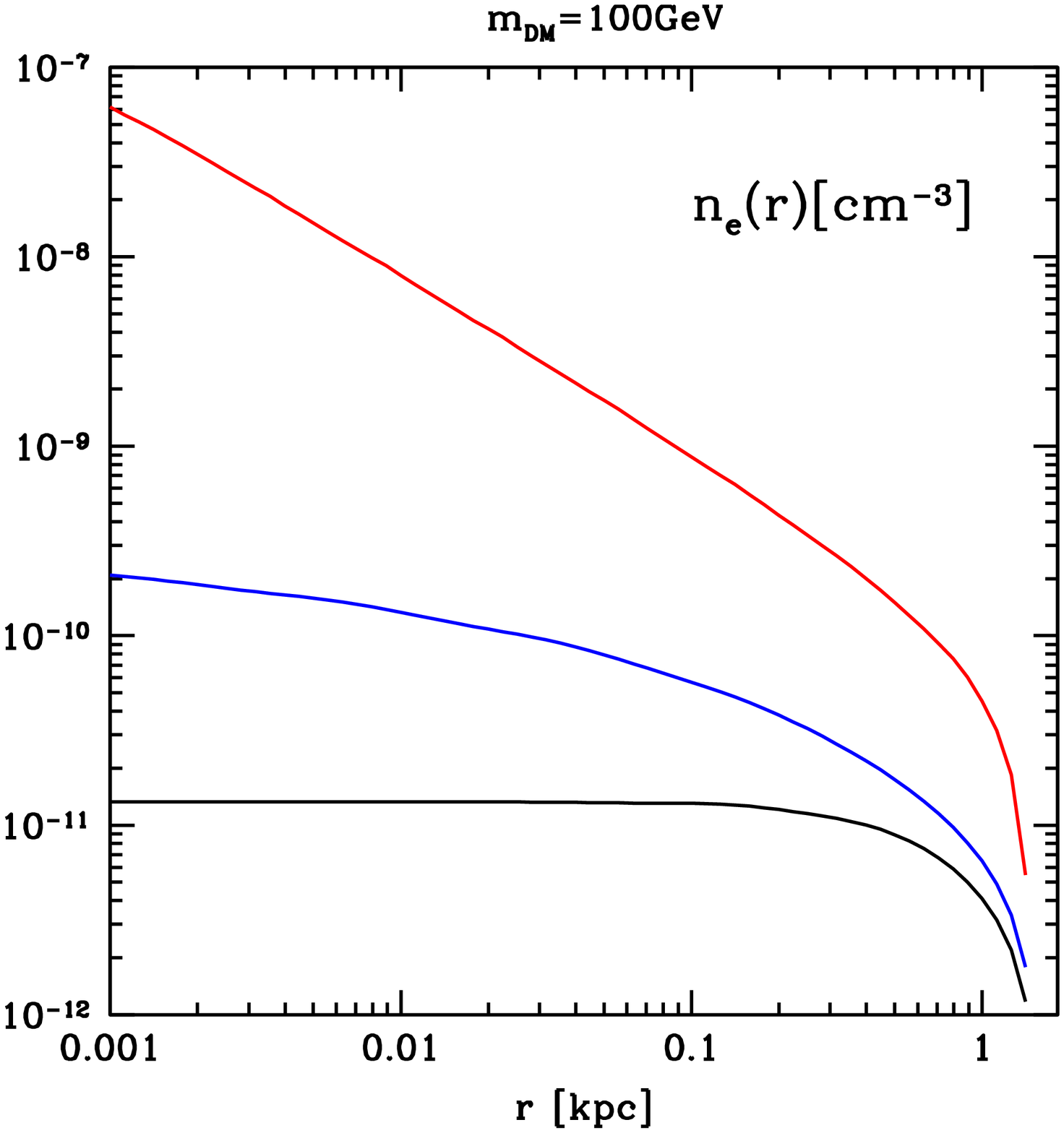}
\caption{The $e^\pm$ density for different halo profiles assumed
The red curve is the result for M99 profile, blue for NFW and
black for B95. \label{WIMPne_halo}} }

In Fig.~\ref{WIMPne_halo} we plot the density distribution of
electrons produced by DM annihilation for different DM halo density
profiles, all of which are obtained with the Green's function
method. As expected, the strongly cuspy M99 profile produces the
largest $e^\pm$ density, followed by NFW profile and then B95
profile. In the center, the density of electrons of M99 profile is
several orders of magnitude higher than that of NFW profile. Thus,
the density profile of DM halo is crucial in determining the
strength of the SZ$_{\rm DM}$ signal.

\subsection{The SZ$_{\rm DM}$ signal}

\FIGURE{
\begin{minipage}[t]{4.0cm}
\includegraphics [scale=0.35]{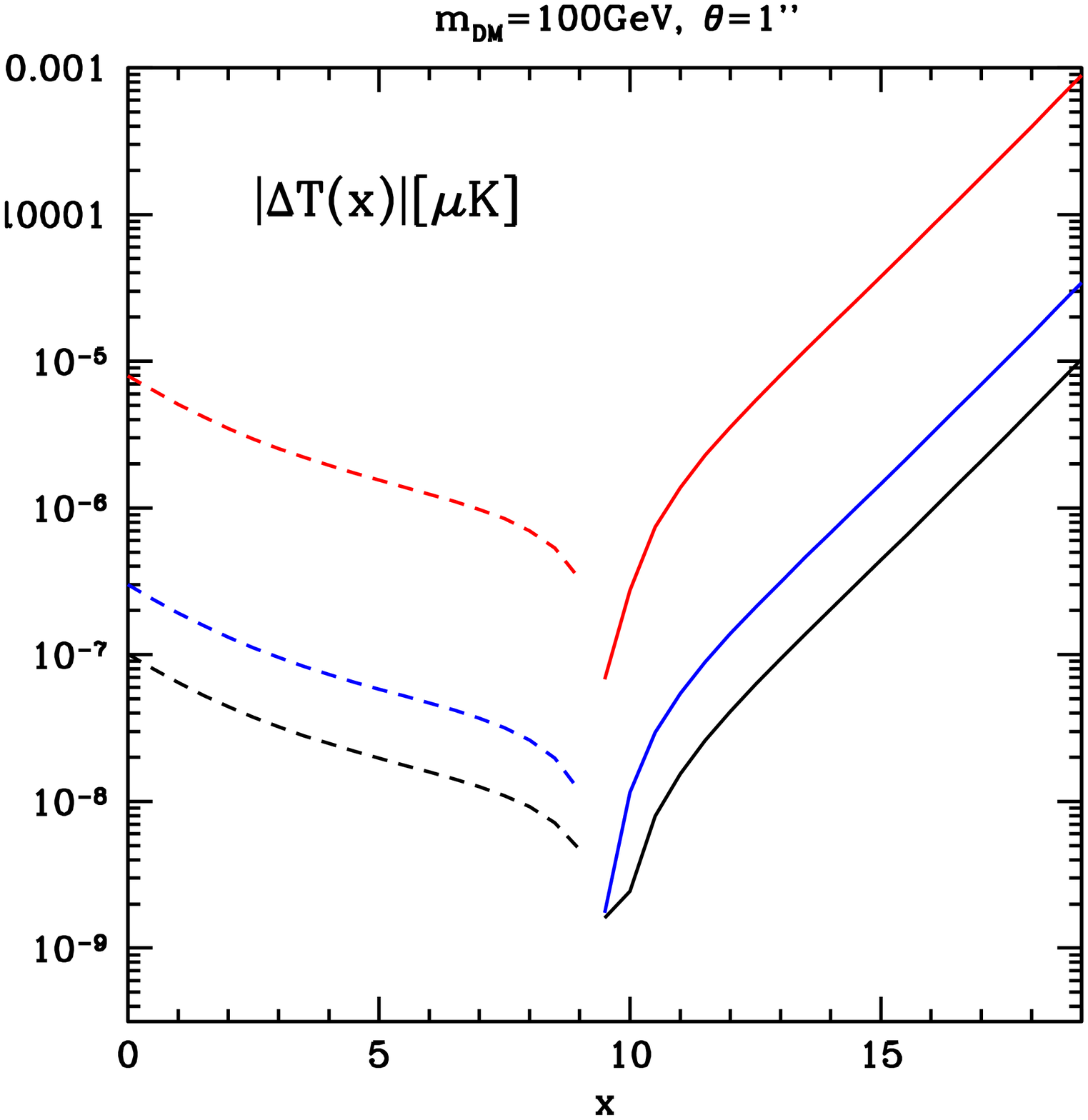}
\end{minipage}
\hfill
\begin{minipage}[t]{7.0cm}
\includegraphics [scale=0.35]{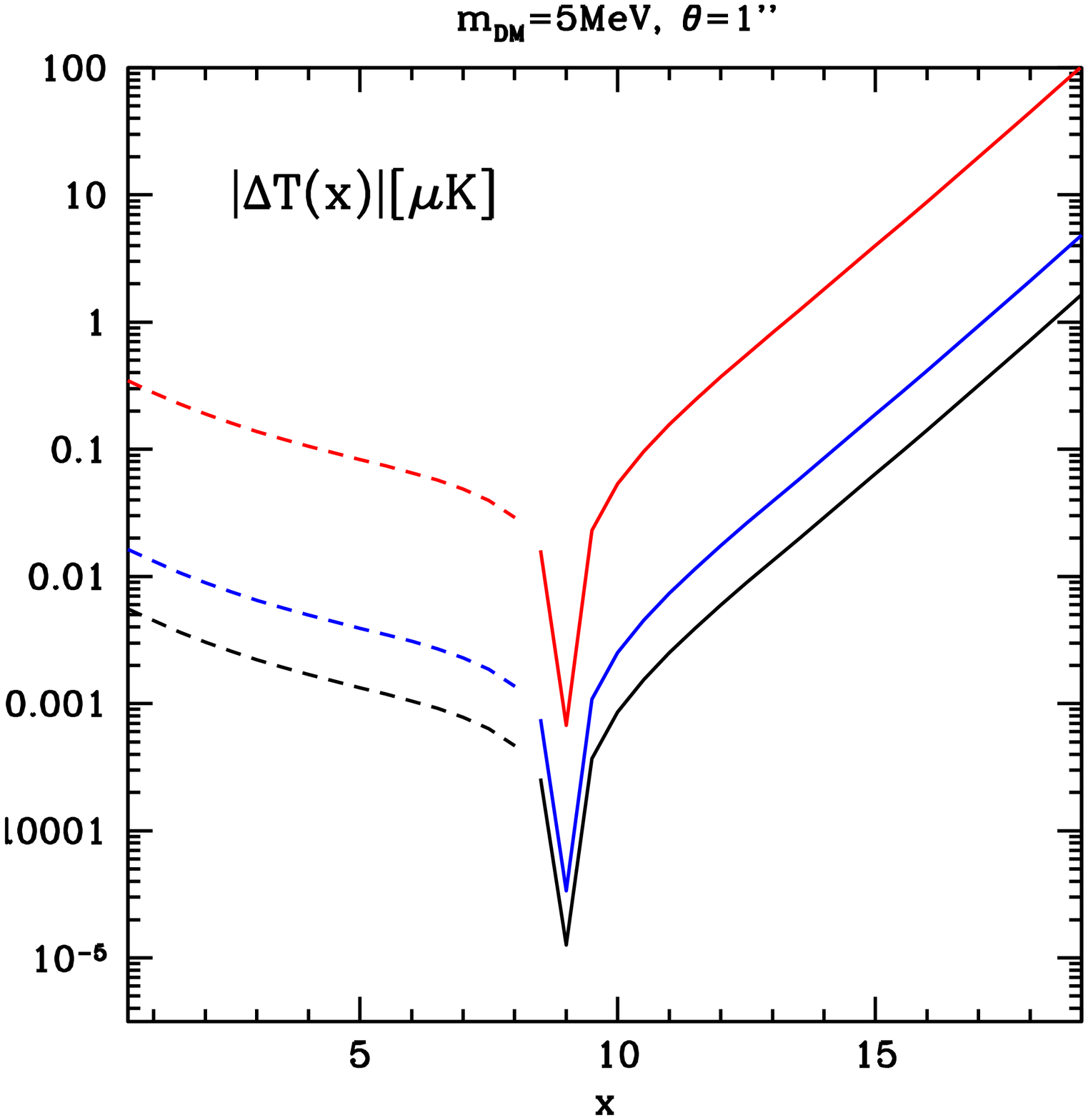}
\end{minipage}
\caption{ The SZ effect induced by the neutralino~(left) and
LDM~(right) annihilations. The result for M99 profile~(red), NFW
profile~(blue) and B95 profile~(black) are shown.
\label{sz_profile}}}

\FIGURE{
\begin{minipage}[t]{4.0cm}
\includegraphics [scale=0.35]{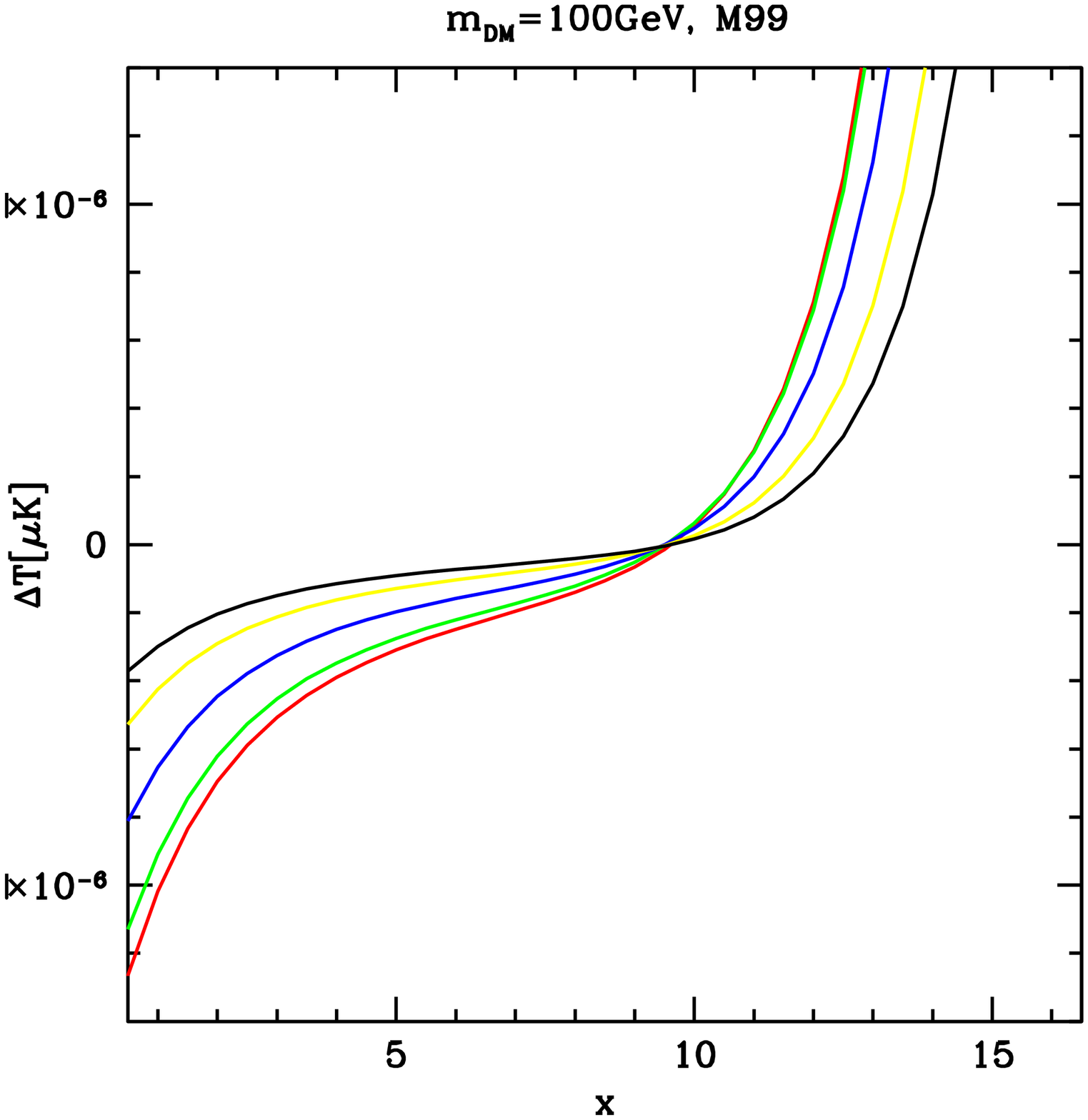}
\end{minipage}
\hfill
\begin{minipage}[t]{7.0cm}
\includegraphics [scale=0.35]{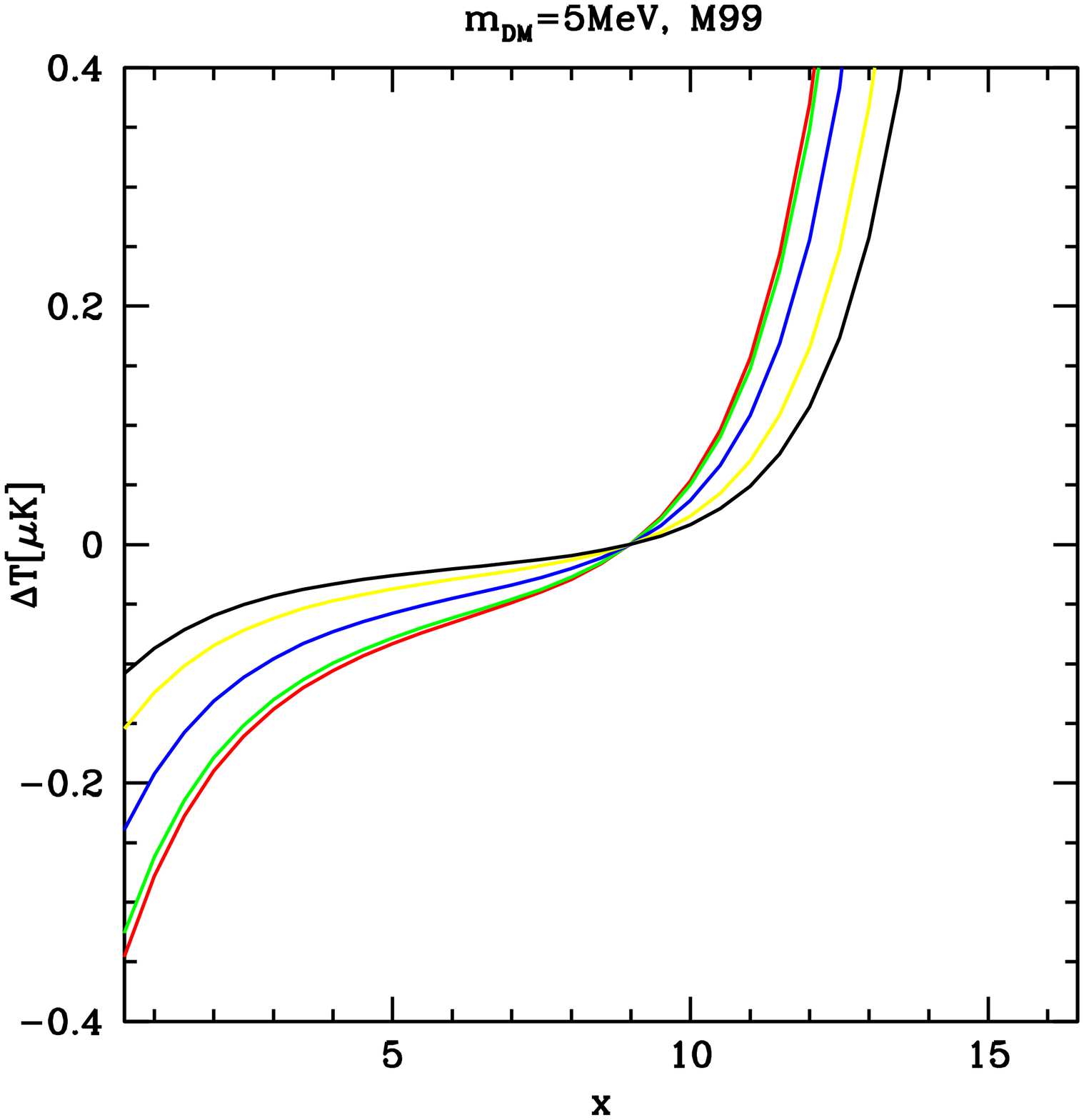}
\end{minipage}
\caption{Spectral Distortion as a function of $x=h\nu/kT_{\rm
CMB}$ for the case of Draco, with beam size ranging from
$1''$~(red), $6''$~(green), $1'$~(blue), $6'$~(yellow) to
$15'$~(black). \label{sz_size}}}

The expected spectral distortion of CMB due to the energetic
$e^{\pm}$ from DM annihilation observed with $1''$ beam aimed at the
center of the Draco dSph are shown in Fig.~\ref{sz_profile}. The
left panel is for a 100 GeV neutralino, and right panel is for a 5 MeV
LDM. The three curves in the figure are for M99, NFW and B95 DM halo
profiles respectively. As expected, the strongly cuspy profile (M99)
yields greater temperature distortion, while NFW and B95 profiles
yield curves which are similar but with less magnitude. We plotted
only the absolute value of the temperature distortion in this
figure, note that at low frequency (small $x$) we have $\Delta T<0$,
while at high frequency (large $x$) $\Delta T>0$. As a result the
magnitude of the temperature distortion appears to drop in the
middle of the figure, where it reverses its sign. Unlike the case of
the thermal SZ effect, this null point is not always fixed at 217
GHz, but varying slightly.

The effects of different beam sizes are illustrated in
Fig.~\ref{sz_size}. The red, green, blue, yellow and black curves
are the observed temperature distortion for the beam size $1''$,
$6''$, $1'$, $6'$ and $15'$ respectively. The smaller beam probes a
smaller region around the center, hence the SZ$_{\rm DM}$ signal is
larger.

\FIGURE{
\includegraphics[scale=0.46]{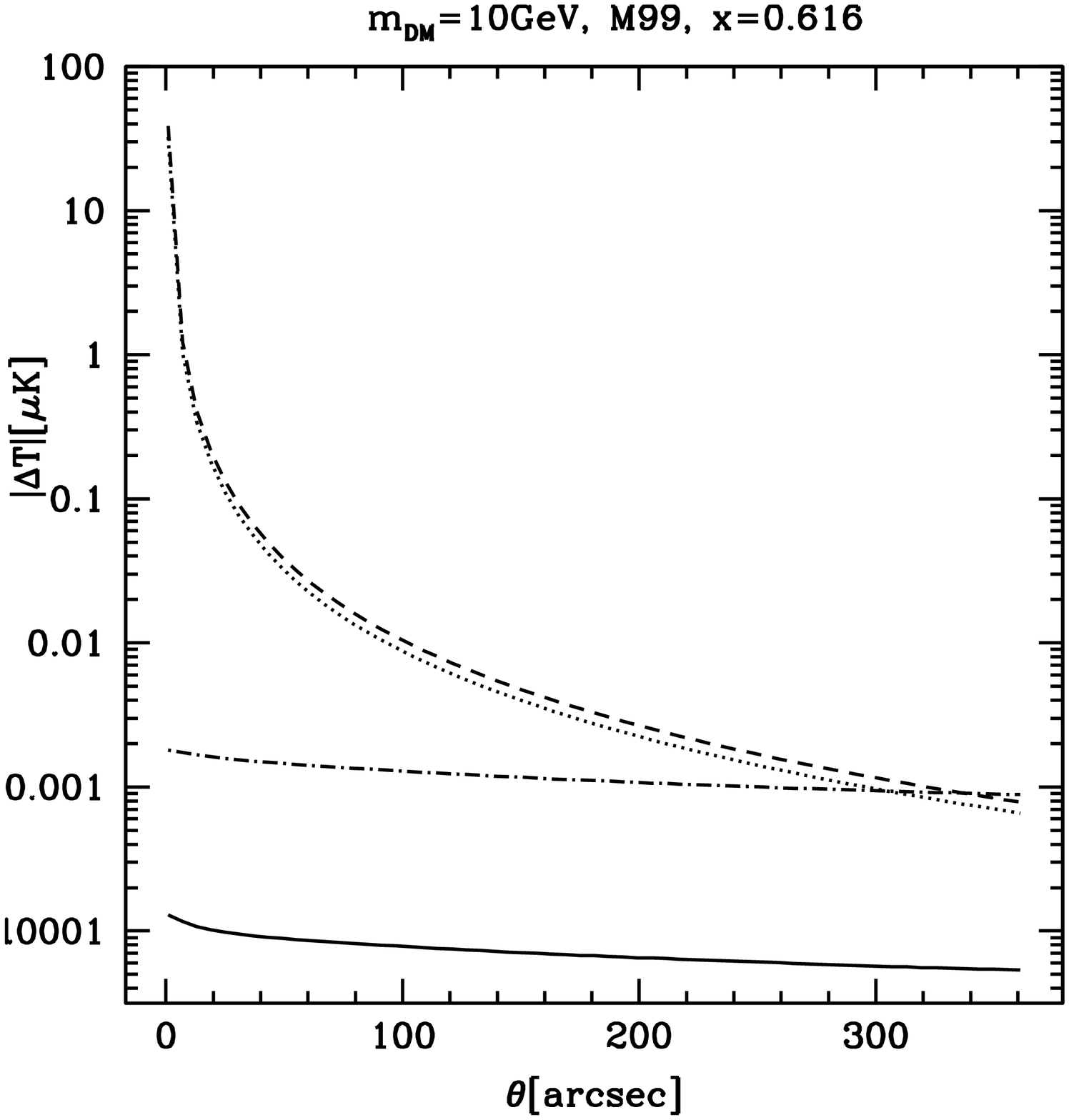}
\caption{SZ$_{\rm DM}$ signal as a function of observing beam
width $\theta$. The solid line is our fiducial model with DarkSUSY
electron spectra. Other lines use analytical spectra, with
different approach to solve the transport equation: Green's
function solution (dot-dashed), ``approximate solution'' (dotted),
and ``no-diffusion solution'' (dashed) respectively.
 \label{varBW}} }

We find that the magnitude of the signal is much smaller than the
previous estimate in the literature (e.g. Ref.\cite{sz_draco}). In
Fig.~\ref{varBW} we plot the SZ$_{\rm DM}$ signal as a function of
the beam width at $x=0.616$~($\nu=35$GHz) which is similar to the
corresponding value in Ref.\cite{sz_draco}. We also take the same
values as $m_{\rm DM}=10\rm GeV$, $<\sigma v>=1.0\cdot 10^{-26}\rm
cm^3~s^{-1}$ in order to have a direct comparison. The solid line
is for DarkSUSY calculated electron injection spectrum with
Green's function solution of the propagation equation. Other lines
use the analytical form of the electron spectra which were shown
as the dashed line in Fig.~\ref{dNdE}. The dot-dashed line in
Fig.~\ref{varBW} is the result from our full solution with the
Green's function, while the dotted line is for the ``approximate
solution'', and the dashed line is for the ``no-diffusion
solution'' respectively. Due to the large difference in the low
energy electrons, the resulting SZ$_{\rm DM}$ signals are quite
different between the DarkSUSY and the analytic spectra. On the
other hand, if one neglects diffusion or only treats it as a
global reduction in the total number of electrons within the halo
as were done in previous works, the signal could be much higher
near the center of the halos. With the diffusion effect included,
however, the brightness temperature only increases slightly near
the center. Only when the beam width become large enough, which is
comparable with the region occupied by the stellar component, the
approximate model can be used. For the 10 GeV neutralino, even for
the M99 profile the expected signal in our fiducial model~(shown
as the solid line in Fig.~\ref{varBW}) at low frequency is well
below $10^{-3}\microK$, which is not observable with the current
or next generation instruments. At the high frequency end, the
signal is higher, but it is difficult to observe from most sites
on the ground. Furthermore, although the dSphs do not have a
thermal SZ background, such small signal might be swamped by other
foreground or background noises.

\FIGURE{
\includegraphics[scale=0.45]{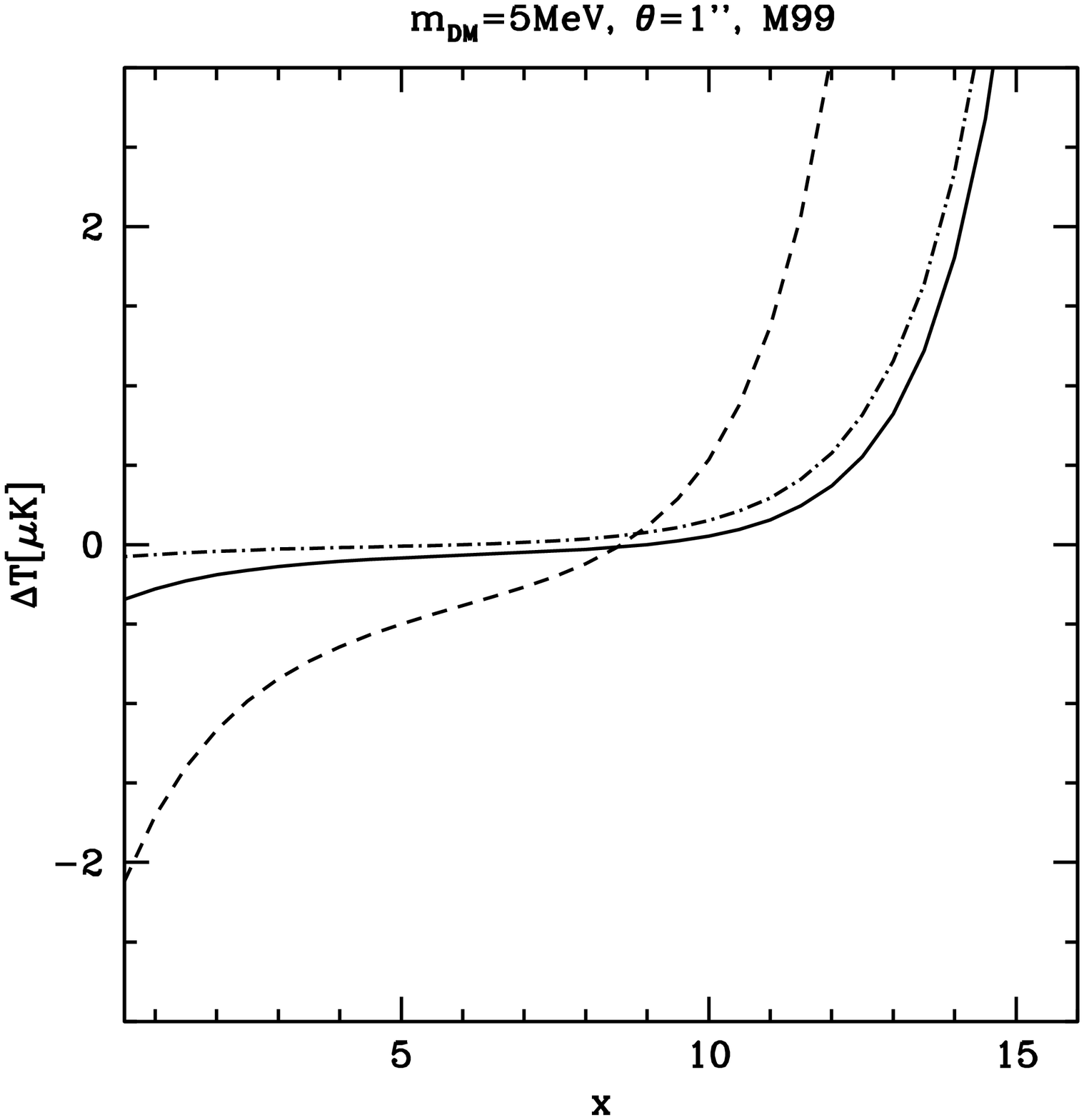}
\caption{The temperature variation for beam size $1''$ induced by
LDM annihilating in dark halo with M99 density profile. The solid
line corresponds to our benchmark value $D_0=3.1\times10^{26} \rm
cm^2~s^{-1}$ and $\delta=1/3$. The dot-dashed line is for the same
$\delta$ but $D_0=1.0\times10^{27} \rm cm^2~s^{-1}$, and the
dashed line is for the same $D_0$ but $\delta = 2/3$.
\label{sz_dif}.} }

Finally, we illustrate in Fig.~\ref{sz_dif} the effect of
different diffusion parameters in $D(E)$ which are not yet
well-constrained. We plot the spectral distortion for three groups
of $D_0$ and $\delta$. The solid line corresponds to our benchmark
value $D_0=3.1\times10^{26} \rm cm^2~s^{-1}$ and $\delta=1/3$. The
dot-dashed line is for the same $\delta$ but $D_0=1.0\times10^{27}
\rm cm^2~s^{-1}$. The dashed line adopts the same $D_0$ but
$\delta = 2/3$. For this purpose we only choose the LDM case to
plot. We see that the SZ$_{\rm DM}$ effect decreases with
increasing $D_{0}$ or decreasing $\delta$. But when the diffusion
parameters vary within a reasonable range, the basic conclusions
remain unchanged.

In the above we have discussed various kinds of effects on SZ$_{\rm
DM}$, taking Draco for example. For other luminous dSphs as listed
in Table \ref{tab:sample}, we present the values of $\Delta T$ at 35
GHz and 1 THz in Table \ref{tab:dSphs1} for the neutralino case, and in
Table \ref{tab:dSphs2} for the LDM case. M99 profile is assumed. The
particle physics parameters of DM are the same as described in Sec.
2.3. Since we have essentially assumed the same density profile for
these dSphs, the magnitudes of their SZ$_{\rm DM}$ effects are
determined mostly by the the mass $\rm M_{\rm vir}$ of each galaxy.
The differences in distance, on the other hand, do not affect the
SZ$_{\rm DM}$ effect. Thus, with these assumptions, dSphs with
larger $\rm M_{\rm vir}$ would generally induce larger temperature
distortion. This can be easily seen by comparing the result for
Draco and Leo I, which have the same $r_h$ but different $\rm M_{\rm
vir}$. Another essential factor in determining the final temperature
distortion is the size of the diffusion zone in each dwarf galaxy.
The value of $r_h$ is generally adopted as twice of the radius of
the stellar component (if this value is larger than the tidal
radius, we set $r_h=r_t$). For smaller $r_h$, a larger fraction of
$e^{\pm}$ would escape away, and the SZ$_{\rm DM}$ effect is also
smaller. This is reflected clearly through the comparison between
Fornax and Leo I.

\begin{table}
\begin{center}
\caption{Neutralino-like DM induced SZ effect (in units of K) for Local
Group luminous dSphs. We assume $m_{\chi}=100~\GeV, \langle \sigma v
\rangle =3.0\times10^{-26}\cm^3\sec^{-1}$, M99 profile, and
$\theta=1''$.} \label{tab:dSphs1}
\begin{tabular}{lccccccc}
\hline \hline \textbf{dSph} & $r_{h} (\kpc)$ &$M_{\rm
vir}(M_{\odot})$&$35\GHz~(x=0.616)$&$1000\GHz~(x=17.46)$
\\ \hline
Ursa Minor& $1.6$ &$4\times10^{9}$&$-5.44\times10^{-12}$&$2.43\times10^{-10}$\\
Draco$^a$ & $1.6$
&$4\times10^{9}$&$-5.39\times10^{-12}$&$2.32\times10^{-10}$
\\  Leo I& $1.6$ &$1\times10^{9}$&$-6.88\times10^{-13}$&$5.37\times10^{-12}$
\\  Fornax & $5.4$ &$1\times10^{9}$&$-5.12\times10^{-13}$&$1.91\times10^{-10}$
\\  Leo II & $1.04$ &$4\times10^{8}$&$-5.45\times10^{-15}$&$3.35\times10^{-11}$
\\  Carina & $1.7$ &$2\times10^{8}$&$-2.37\times10^{-13}$&$8.97\times10^{-13}$
\\  Sculptor & $3.26$ &$1\times10^{9}$&$-2.61\times10^{-12}$&$9.72\times10^{-11}$
\\  Sextans$^b$ & $4.8$
&$3\times10^{8}$&$-1.61\times10^{-12}$&$5.89\times10^{-11}$\\
\hline\hline
\end{tabular}
\end{center}
$^a$$r_{\rm stellar}$ is around 0.93 kpc, which give a slightly
large $r_h$ as 1.86 kpc. In the calculation, we just use the
$r_h=1.6$~kpc as the diffusion zone for consistency with previous
results.\\
$^b$$r_{\rm stealer}$ is around 4 kpc, which give a really large
$r_h$ as 8 kpc. In the real calculation, we use $r_t$ as the
diffusion zone.
\end{table}

\begin{table}
\begin{center}
\caption{LDM induced SZ effect (in units of K) for Local Group
luminous dSphs. We assume $m_{\rm LDM}=5~\MeV, \langle \sigma v
\rangle =2.5\times10^{-29}\cm^3\sec^{-1}$, M99 profile, and
$\theta=1''$.} \label{tab:dSphs2}
\begin{tabular}{lccccccc}
\hline \hline \textbf{dSph} & $r_{h} (\kpc)$ &$M_{\rm
vir}(M_{\odot})$&$35\GHz~(x=0.616)$&$1000\GHz~(x=17.46)$
\\ \hline
Ursa Minor& $1.6$ &$4\times10^{9}$&$-3.11\times10^{-7}$&$2.68\times10^{-5}$\\
Draco & $1.6$
&$4\times10^{9}$&$-3.11\times10^{-7}$&$2.67\times10^{-5}$
\\  Leo I& $1.6$ &$1\times10^{9}$&$-6.76\times10^{-8}$&$1.48\times10^{-6}$
\\  Fornax & $5.4$ &$1\times10^{9}$&$-2.37\times10^{-7}$&$3.80\times10^{-5}$
\\  Leo II & $1.04$ &$4\times10^{8}$&$-1.63\times10^{-8}$&$4.76\times10^{-6}$
\\  Carina & $1.7$ &$2\times10^{8}$&$-1.97\times10^{-8}$&$1.71\times10^{-6}$
\\  Sculptor & $3.26$ &$1\times10^{9}$&$-1.37\times10^{-7}$&$1.48\times10^{-5}$
\\  Sextans & $4.8$ &$3\times10^{8}$&$-7.76\times10^{-8}$&$1.09\times10^{-5}$
\\
\hline \hline
\end{tabular}
\end{center}
\end{table}

Recently, the PAMELA~\cite{PAMELA}, ATIC~\cite{ATIC} and
FERMI~\cite{FERMI} observations of the cosmic ray electron and
positron energy spectrum indicate a possible excess or even peak in
the cosmic ray electrons and positrons. If this excess is due to DM
annihilations, then the annihilation cross section must be
``boosted'' by a large factor compared with the usual expectation.
Such a boost could be due to, e.g.,  ``Sommerfeld enhancement''
\cite{2009NuPhB.813....1C,2009PhRvD..79a5014A,2009PhLB..671..391P,
2008JHEP...07..058M,2008arXiv0812.0559M}. However, in such a case
the mass of DM particle must be of TeV scale, so even with a 1000-fold
enhancement, the induced SZ$_{\rm DM}$ effect, according to our
estimate, would still be quite small. The DM substructures inside
the dark halo are also thought to be able to ``boost'' the
annihilation signal to some extent \cite{2007JCAP...05..001Y,
2008A&A...479..427L}. However, the substructures near the center of
the halo will be destroyed by tidal force. The ``boost'' is more
effective at large radii \cite{2006NuPhB.741...83B}. Furthermore,
for the cuspy profiles the central density is high enough to
dominate the contribution over that from substructures. Hence the
``boost'', regardless of its origin, would have little impact on our
present calculation.

\section{Conclusion}

In this paper, we calculate the non-thermal SZ effect induced by the
energetic electrons and positrons produced by the annihilation of
neutralino and LDM in dSphs. We take the Draco dwarf as an example to
present most of our results, but we have also applied the same
calculation to other luminous dSphs in the Local Group.

In our calculation, we obtain the equilibrium $e^{\pm}$
distribution by fully solving the diffusion equation with the
Green's function method. We take synchrotron emission, ICS,
ionization loss as well as diffusion loss of $e^{\pm}$ into
account. We find that for small scale systems such as the dwarf
galaxies, diffusion effect is crucial. The $e^{\pm}$ distribution
is much less steep than the source function which is proportional
to $\rho_{\rm DM}^2 $. This important effect was mentioned in
\cite{draco_multi} but was neglected in the numerical calculations
in their previous analysis \cite{sz_draco}, and the resulting
$e^{\pm}$ density could be different by several orders of
magnitude. Another difference is that we considered contributions
from $e^{\pm}$ with relatively low energy (i.e. we do not use any
arbitrary energy cut off). Such low energy $e^{\pm}$ could still
contribute significantly to the SZ$_{\rm DM}$ signal.

The SZ effect by the non-thermal $e^{\pm}$ depends on the density
profile of the DM halo. We considered M99, NFW and B95 profiles. As
expected, for the cuspy M99 profile the effect is much more
significant. Nevertheless, for all profiles we find much smaller
signals than previously claimed. Due to the diffusion effect, even
for the strongly cuspy M99 profile, the SZ effect predicted for neutralino
is too small to be observed with the current or coming generation of
instruments. The possible astrophysical foregrounds or
contamination may even make the detection more difficult.

Considering that a smaller mass of DM will result in a higher DM
number density and annihilation rate, we also investigated the case
of LDM, i.e. DM candidate with MeV mass scale. In this case, there
is some hope of detecting the DM induced SZ effect, though the
effect is also much smaller than previous claims. For the strongly
cusped profile assumed, $\Delta T$ could reach tens of $\mu$K when
the frequency is around or larger than THz. However, we note that
THz observation is difficult to do on ground, with the possible
exception of a few sites in the Antarctica high plateau.

One of the largest uncertainty in our calculation is the diffusion
coefficient of the dSphs, which depends on the magnetic field strength,
for which we have no data. We made an assumption on its form and
value for the present calculation. It is possible that the
magnetic field is much weaker than we assumed, in that case, the
SZ effect by non-thermal $e^{\pm}$ would be even smaller.

These results show that the non-thermal SZ effect induced by DM
annihilation in dSphs is small and difficult to observe, even though
the expected astrophysical contamination is small. Combined with our
earlier results for clusters (Paper I), we conclude that the
non-thermal SZ$_{\rm DM}$ effect is perhaps not a powerful method to
detect or constrain DM annihilations.

\acknowledgments We thank Pengjie Zhang, Bin Yue, and the 
anonymous referee for helpful discussion 
and suggestions. This work is supported by the National
Science Foundation of China under grants 10773011, 10525314,
10533010, by the Chinese Academy of Sciences under grant No.
KJCX3-SYW-N2, and by the Ministry of Science and Technology
National Basic Science Program (Project 973) under grants
2007CB815401 and 2010CB833000, and by the National Basic Research
Program of China under Grant No. 2009CB824800.

\bibliography{sz}

\bibliographystyle{JHEP}

\end{document}